\documentclass{aa}  

\usepackage{graphicx}
\usepackage{txfonts}
\usepackage[colorlinks=true,
    citecolor=blue]{hyperref}
\begin{document}

   \title{Alfv\'en wave propagation from the photosphere to the corona:  temporal evolution against stationary results}

\titlerunning{Alfv\'en wave transmission from the photosphere to the corona}
\authorrunning{R. Soler}

     \author{Roberto Soler\inst{1,2}}

   \institute{Departament de Física, Universitat de les Illes Balears, 07122 Palma de Mallorca, Spain\\
            \and Institut d’Aplicacions Computacionals de Codi Comunitari (IAC3), Universitat de les Illes Balears, 07122 Palma de Mallorca,
            Spain\\ 
             \email{roberto.soler@uib.es}}

   \date{Received XXX; accepted XXX}

  \abstract{Recent observations have confirmed that a significant fraction of the coronal Alfv\'enic wave spectrum originates in the photosphere. These waves travel from the photosphere to the corona, overcoming the  barriers of reflection and dissipation posed by the chromosphere. Previous studies have theoretically calculated the chromospheric reflection, transmission, and absorption coefficients for pure Alfv\'en waves under the assumption of stationary propagation. Here, we relax that assumption and investigate the time-dependent propagation of Alfv\'en waves driven at the photosphere. Using an idealized chromospheric background model, we compare the  coefficents  obtained from time-dependent simulations with those derived under the stationary approximation. Additionally, we examine the impact of the spatial resolution in the numerical simulations.  Considering a spatial resolution of 250~m, we find that the time-dependent transmission coefficient converges  to the stationary value across the entire frequency range after only a few chromospheric Alfv\'en crossing times, while the reflectivity displays a good convergence for frequencies lower than 30 mHz. The absorption coefficient also converges  for wave frequencies above  1~mHz, for which chromospheric dissipation is significant. In contrast, at lower  frequencies, wave energy dissipation is weak and  the time-dependent simulations typically overestimate the absorption. Inadequate spatial resolution artificially enhances the chromospheric reflectivity,  reduces wave transmission to the corona, and poorly describes the wave energy absorption.  Overall, the differences between the stationary and time-dependent approaches are only minor and gradually decrease as spatial resolution and simulation time increase, which reinforces the validity of the stationary approximation. }


   \maketitle
%

\section{Introduction}

Recent high-resolution and high-cadence observations \citep[see][]{morton2025a,morton2025} have confirmed that the coronal Alfv\'enic power spectrum is dominated by waves excited in the photosphere. Previous observational and theoretical studies demonstrated that horizontal convective motions in the photosphere can efficiently excite nearly incompressible  Alfv\'enic waves \citep[see, e.g.,][]{Choudhuri1993,Noble2003,Chitta2012,morton2013,VanKooten2024}. These waves then propagate from the photosphere to the corona through the chromosphere and transition region, overcoming the strong reflection and dissipation  associated, respectively, with  steep vertical gradients and partial ionization effects \citep[see][]{soler2017,soler2019}.

Several previous studies have theoretically investigated the propagation of Alfv\'en waves through the stratified chromosphere and their damping due to partial ionization \citep[see, e.g.,][among others]{depontieu2001,leake2005, goodman2011, tu2013, arber2016,Pelekhata2021,Kraskiewicz2023}. Assuming the stationary state, \citet{soler2019} computed the intrinsic reflection, transmission, and absorption coefficients for torsional Alfv\'en waves propagating from the photosphere to the corona. They showed that the transmission of waves to the corona depends on the competition between strong reflection at low frequencies and efficient dissipation at high frequencies. As a result of this interplay, the transmission coefficient peaks at a frequency around 5~mHz, although the exact peak frequency depends on the photospheric magnetic field strength. Despite the strong chromospheric filtering, about 1\% of the wave energy driven in the photosphere is able to reach coronal heights. Nevertheless, this modest fraction may represent a significant contribution to the coronal power spectrum \citep{morton2025}.

The results of \citet{soler2019} were obtained under the stationary approximation. The aim of this paper is to relax that assumption and consider the full temporal evolution of Alfv\'en waves propagating from the photosphere to the corona. Our goal is to use the time-dependent results to compute the reflection, transmission, and absorption coefficients, and to compare them with their stationary counterparts as a way to test the validity of the stationary approximation. Since we intend to explore a broad frequency range and perform a large number of time-dependent simulations, we adopt a simplified 1D version of the model by \citet{soler2019}, in which the expansion of the magnetic field with height is neglected. Nevertheless, the background model retains the effects of gravitational stratification and partial ionization, ensuring that the transmission profile still incorporates the essential physics of the more complete model. The consideration of multi-dimensional models in this context  is left for forthcoming works.

This paper is organized as follows. Section~\ref{sec:model} contains the description of the background model and the basic equations. Then, results in the stationary approximation are given in Section~\ref{sec:sta}, while those of the full temporal evolution are presented and analyzed in Section~\ref{sec:temp}. Finally, Section~\ref{sec:conc} contains some concluding remarks.

\section{Model and basic equations}
\label{sec:model}

The solar chromosphere is a highly dynamic environment. When studying wave propagation in this region, we encounter the inherent complexity of distinguishing wave activity from the natural evolution of the medium in which these waves propagate. In this paper, we follow the same approach as in our previous works \citep{soler2017,soler2019} and adopt a simplified version of the chromosphere made of a static model. While this model provides a limited representation of the chromosphere that does not allow to study all the chromospheric dynamics, it allows for a detailed analysis of wave behavior and contains the basic ingredients that affect Alfv\'en wave propagation in the partially ionized chromospheric plasma.

\begin{figure}
   \centering
   \includegraphics[width=0.95\columnwidth]{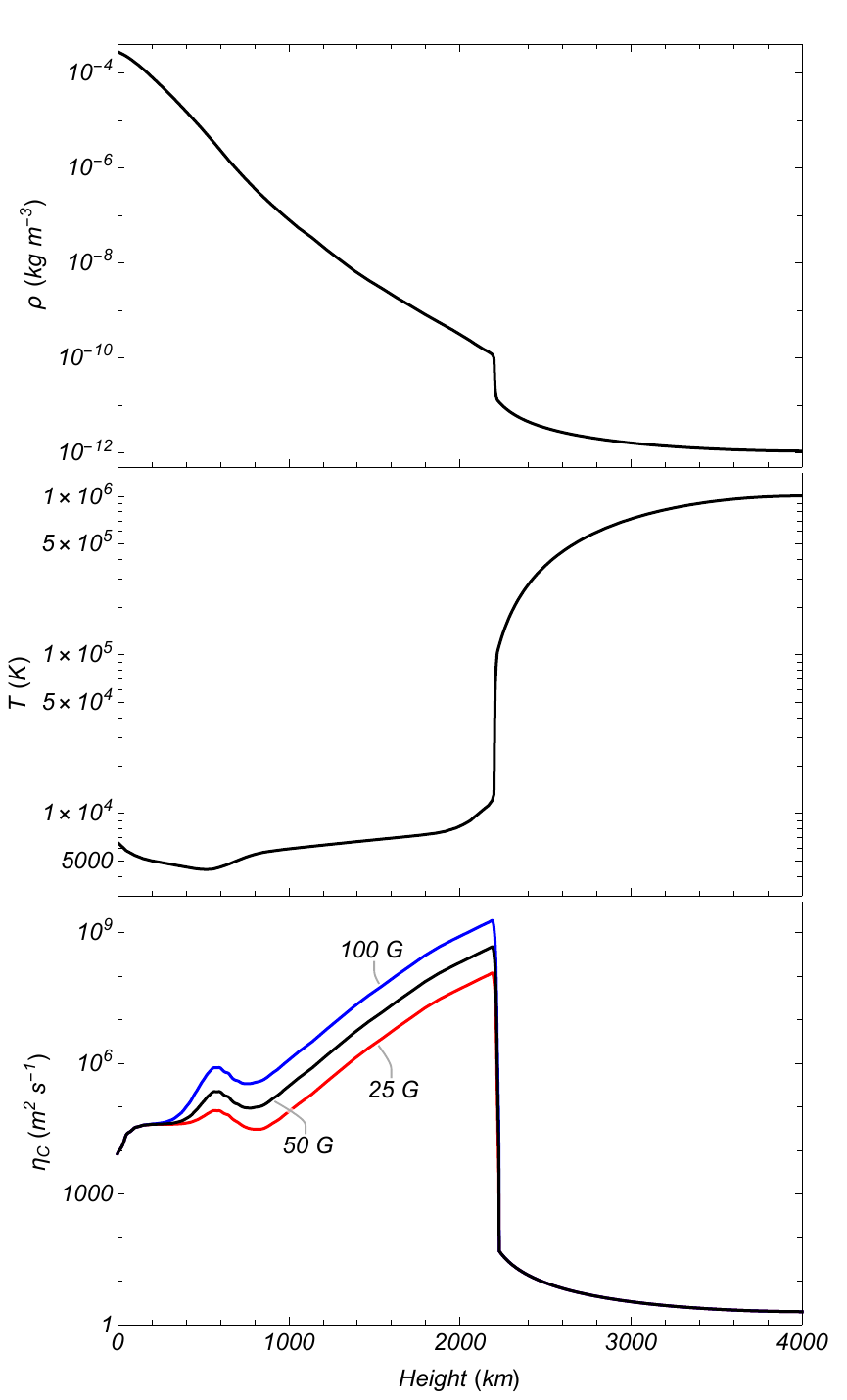}
      \caption{Background atmospheric model. Dependence with height over the photosphere of  density (top), temperature (mid), and Cowling's coefficient (bottom). Three different values of $B_0$ are considered in the bottom panel.}
         \label{fig:model}
   \end{figure}

We consider a static, plane-parallel, and gravitationally stratified medium to represent the lower solar atmosphere. The physical properties of the plasma are based on the semi-empirical model C from \citet{Fontenla1993}, hereafter FALC model, representing an average quiet-Sun scenario. We use the FALC model because it provides the height-dependent variations of neutral and ionized helium densities, in addition to hydrogen densities,  which are important for calculating the plasma dissipative coefficients. This information is not available in more recent tabulated models. The FALC model, originally a chromospheric model, has been extended to include the lower corona. The vertical coordinate, denoted as $z$, is aligned with the vertical coordinate of FALC. Thus, the model extends from the top of the photosphere, $z=0$,  through the chromosphere and the transition region (located at approximately $z \approx 2200$~km), up to the lower corona at $z = L = 4000$~km. The variations of density, $\rho$, and temperature, $T$, with height can be seen in  Figure~\ref{fig:model}. The thin transition region is visible as a sudden, steep variation of both density and temperature.

For simplicity, in this work we consider a straight, vertical magnetic field, namely ${\bf B} = B_0 \hat{e}_z$, with $B_0$ constant. The effects of field line expansion were analyzed at length in \citet{soler2017,soler2019}. It was found that the expansion of the magnetic field  with height favors Alfv\'en wave transmission to the corona, but it also enhances the dissipation in the lower chromosphere owing to the development of cross-field gradients by the phase-mixing process. While the consideration of magnetic field expansion is important for an accurate determination of wave transmission through the lower atmosphere, the goal of this work is to compare stationary and time-dependent results. For this purpose and to keep the execution times within reasonable limits, it suffices to use a simpler straight magnetic field. Therefore, we conveniently reduce the problem of Alfv\'en wave propagation to a 1.5D problem and, in this context, the value of $B_0$ should be understood as an average value of the magnetic field strength. A similar approach was followed in, e.g., \citet{arber2016}. 

We note that since the adopted magnetic field is very simple, the waves studied here are pure Alfv\'en waves whose character does not change as they propagate from the photosphere, where the waves are driven, to the corona. In more complex environments, Alfv\'en waves can also be generated by mode conversion from fast and slow magnetosonic waves, but this is not possible in this model \citep[see, e.g.,][]{cally2011,cally2022}.  

Owing to the high values of the ion-neutral collision frequency in the lower atmosphere \citep[see, e.g.,][]{ballester2018,soler2024}, we restrict ourselves to using the single-fluid MHD approximation.  The linearized equations for the discussion of Alfv\'en waves in the model are,
\begin{eqnarray}
    \frac{\partial v_\perp}{\partial t} &=& \frac{B_0}{\mu \rho}  \frac{\partial b_\perp}{\partial z}, \label{eq:main1} \\
     \frac{\partial b_\perp}{\partial t} &=& B_0 \frac{\partial v_\perp}{\partial z} + \frac{\partial}{\partial z} \left( \eta_{\rm C} \frac{\partial b_\perp}{\partial z} \right), \label{eq:main2}
\end{eqnarray}
where $v_\perp$ and $b_\perp$ are the perpendicular components of the velocity and magnetic field linear perturbations, respectively, associated to the Alfv\'en waves, $\mu$ is the magnetic permeability, and $\eta_{\rm C}$ is the coefficient of Cowling's (or total) magnetic diffusion given by,
\begin{equation}
    \eta_{\rm C} = \eta + B_0^2 \eta_{\rm A}, \label{eq:cowling}
\end{equation}
where $\eta$ and $\eta_{\rm A}$ are the Ohmic and ambipolar coefficients, respectively. Expressions to compute these coefficients as functions of the local plasma properties are given in, e.g., \citet{Khomenko2012,soler2015}. Equation~(\ref{eq:cowling}) shows that the relative weight of the two effects on the total diffusion coefficient depends upon the value of $B_0$.  Figure~\ref{fig:model}  displays the variation of $\eta_{\rm C}$ with height  for various values of $B_0$. In the very low chromosphere, as well as above the transition region, the magnetic field strength is irrelevant for the value of $\eta_{\rm C}$, which indicates that Ohmic diffusion dominates in those regions. Indeed,  the plasma becomes fully ionized above the transition region, so that ambipolar diffusion is absent and $\eta_{\rm C} = \eta$. Conversely, ambipolar diffusion is important in the middle and upper chromosphere.

\section{Stationary results}
\label{sec:sta}

In the stationary state of wave propagation, we assume that the temporal dependence of the perturbations is given by $\exp \left( - i\omega t \right)$, with $\omega$ the angular frequency in rad~s$^{-1}$. The linear frequency is $f=\omega/2\pi$ and is expressed in Hz. With this temporal prescription, Equations~(\ref{eq:main1}) and (\ref{eq:main2}) can easily be combined to arrive at a differential equation involving $b_\perp$ alone, namely
\begin{equation}
     \frac{\partial}{\partial z} \left( \left( v_{\rm A}^2 - i \omega \eta_{\rm C} \right) \frac{\partial b_\perp}{\partial z} \right) + \omega^2 b_\perp = 0, \label{eq:stationary}
\end{equation}
where $v_{\rm A} = B_0 /\sqrt{\mu\rho}$ is the Alfv\'en speed. For a given frequency, Equation~(\ref{eq:stationary}) can be integrated between $z=0$ (the photosphere) and $z=L$ (the corona) as a two-point boundary-value problem, following the same method as in \citet{soler2017,soler2019}. A short summary is provided here.

We resort to the so-called Els\"asser variables \citep{Elsasser1950} to set physically meaningful boundary conditions. We define as $Z^\uparrow$ and $Z^\downarrow$ the Els\"asser variables that represent upward-propagating and downward-propagating Alfv\'en waves, respectively, namely
\begin{equation}
      Z^{\uparrow} = v_{\perp} - \frac{b_{\perp}}{\sqrt{\mu\rho}}, \qquad
    Z^{\downarrow} = v_{\perp} + \frac{b_{\perp}}{\sqrt{\mu\rho}}. 
\end{equation}
We assume that the waves are driven at the photosphere. Then, at the lower photospheric boundary, there would be a superposition of both upward-propagating and downward-propagating waves,  so that $Z^\uparrow$ and $Z^\downarrow$ would be both nonzero at that boundary. The upward-propagating waves would be the ones generated by the driver itself and the downward-propagating waves would correspond to the waves reflected from the above chromosphere \citep[see, e.g.,][]{soler2025}. Regardless of what combination of $Z^\uparrow$ and $Z^\downarrow$ happens at this boundary, we can simply assign an arbitrary amplitude to $b_\perp$ at $z=0$. Conversely, the upper coronal boundary is considered to be perfectly transparent to the waves incident from below and there are no waves coming from the above corona, so that $Z^{\uparrow}\neq 0$ and $Z^{\downarrow} = 0$ at $z=L$. These considerations result in the  boundary conditions for $b_\perp$, namely
\begin{eqnarray}
    b_\perp &=& 1,\quad \text{at}\quad z=0, \label{eq:bound1}\\
    \frac{\partial b_\perp}{\partial z} &=& \frac{i\omega}{v_{\rm A,c}} b_\perp,\quad \text{at}\quad z=L,  \label{eq:bound2}
\end{eqnarray}
where $v_{\rm A,c}$ denotes the Alfv\'en speed at the coronal boundary.

The energy flux carried by the Alfv\'en waves  is,
\begin{equation}
    F = -\frac{B_0}{\mu}v_\perp b_\perp.
\end{equation}
 We can decompose the energy flux as $F = F^\uparrow - F^\downarrow$, where $F^\uparrow$ and $F^\downarrow$ are the upward and downward fluxes, respectively. Using again the Els\"asser variables, we can write
\begin{equation}
    F^\uparrow = \frac{B_0}{4} \sqrt{\frac{\rho}{\mu}}Z^{\uparrow}Z^{\uparrow},\qquad F^\downarrow = \frac{B_0}{4} \sqrt{\frac{\rho}{\mu}}Z^{\downarrow}Z^{\downarrow}. \label{eq:fluxes}
\end{equation}
We average these expressions over one wave period to compute the net fluxes \citep[see, e.g.,][]{walker2005}. Then,  we use the net fluxes at the boundaries to compute the reflectivity, $\mathcal R$, and transmissivity, $\mathcal T$, as 
\begin{eqnarray}
    \mathcal{R} = \frac{\left< F^\downarrow \right>_{z=0}}{\left<F^\uparrow \right>_{z=0}}, \qquad \mathcal{T} = \frac{\left<F^\uparrow \right>_{z=L}}{\left<F^\uparrow \right>_{z=0}}, \label{eq:reftran}
\end{eqnarray}
where we used the notation $\left< \cdots \right>$ to denote the temporal averaging of the fluxes. The coefficients $\mathcal R$ and $\mathcal T$ represent the fractions of the driven wave energy that are reflected back to the photosphere and transmitted all the way to the corona, respectively. However, the chromosphere is a dissipative medium, with dissipation being represented in this model by Cowling's diffusion. Thus, a fraction of the energy is also absorbed. By invoking energy conservation, we compute the absorption, $ \mathcal A$, as, 
\begin{equation}
    \mathcal{A} = 1 - \mathcal{R} - \mathcal{T}. \label{eq:abs}
\end{equation}

 \begin{figure}
   \centering
   \includegraphics[width=0.95\columnwidth]{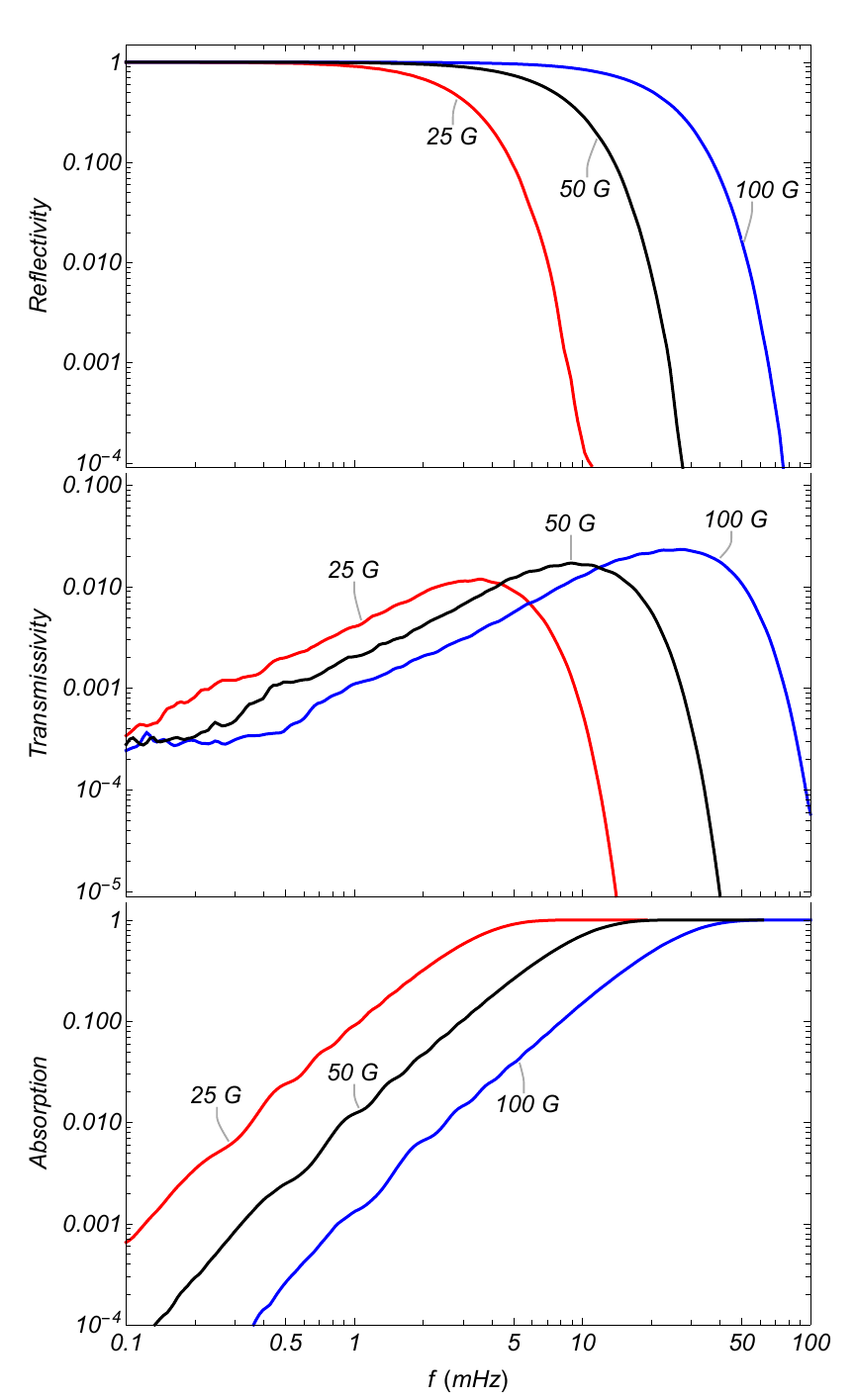}
      \caption{Stationary results. Reflectivity (top), transmissivity (mid), and absorption (bottom) as functions of the Alfv\'en wave frequency. Three different values of $B_0$ are considered.}
         \label{fig:stationary}
   \end{figure}

We have numerically integrated Equation~(\ref{eq:stationary})  with the boundary conditions of Equations~(\ref{eq:bound1})--(\ref{eq:bound2}) for wave frequencies ranging from 0.1~mHz to 100~mHz.  Later, we have computed the coefficients $\mathcal R$, $\mathcal T$, and $\mathcal A$ as functions of the frequency. These results are displayed in Figure~\ref{fig:stationary} for $B_0=$~25, 50, and 100~G. It is found that the chromosphere is an efficient filter for the Alfv\'en waves. Essentially, reflection dominates for low frequencies and dissipation (absorption) does so for high frequencies. This competition between reflection and dissipation results in a transmission coefficient that displays a maximum for intermediate frequencies, although the  energy percentage that reaches the corona remains very small in the whole frequency range. As shown in \citet{soler2019}, the transmission coefficient can be well approximated by a skewed log-normal distribution, which has a steep tail towards the absorption-dominated high frequencies and a more gentle tail towards the reflection-dominated low frequencies.

The magnetic field strength determines both  frequency location and  value of the transmissivity maximum. As $B_0$ increases/decreases, the maximum shifts towards higher/lower frequencies and its amplitude slightly increases/decreases. This happens because changing $B_0$ modifies the relative roles of reflection and dissipation. Reflection gains/loses importance when $B_0$ increases/decreases, and the opposite occurs to the absorption. This is related to the dependence of the wavelength with the magnetic field strength (the wavelength is proportional to $B_0$ for an Alfv\'en wave). The longer the wavelength, the stronger the reflection in the stratified chromosphere and in the sharp transition region, while the less efficient the dissipation \citep[see a detailed discussion in][]{soler2024}.

 \begin{figure*}
   \centering
   \includegraphics[width=1.8\columnwidth]{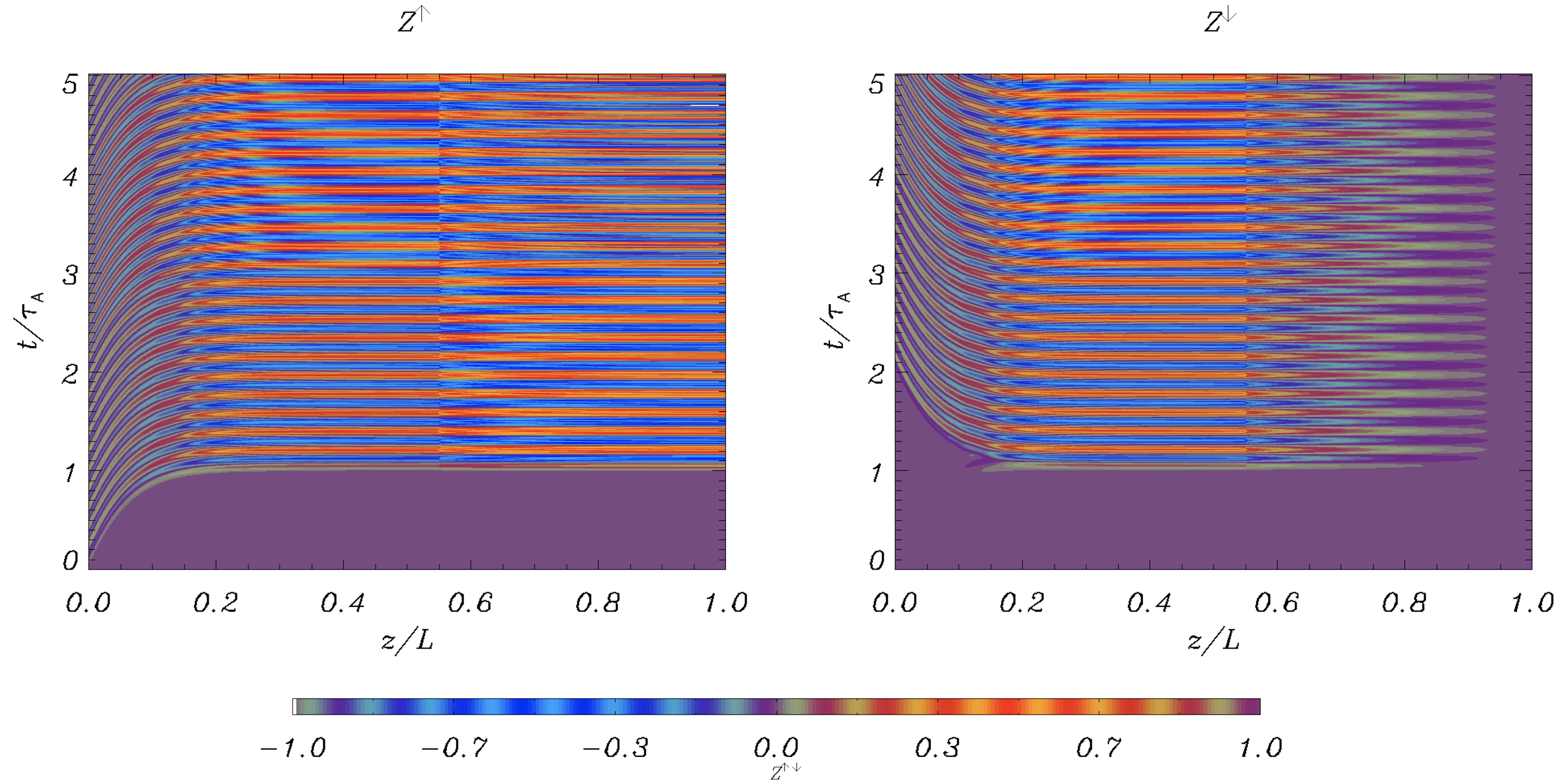}
      \caption{Time-distance diagrams of the Els\"asser variables $Z^\uparrow$ (left) and $Z^\downarrow$ (right) for an Alfv\'en wave with a frequency of 5~mHz. The amplitudes of the Els\"asser variables are expressed in arbitrary units. Time and distance are normalized with respect to the Alfv\'en crossing time and domain length, respectively.}
         \label{fig:temporal5mHz}
   \end{figure*}

\section{Temporal evolution}
\label{sec:temp}

Now, the goal is to compare the stationary results with those obtained from the full temporal evolution. Here, for simplicity, we  set the magnetic field strength to $B_0 = 50$~G. The stationary case suggests that changing $B_0$ would essentially produce a frequency shift of the transmission curve.

Equations~(\ref{eq:main1}) and (\ref{eq:main2}) form a system of advection-diffusion equations. We have written a  numerical code in which the system is solved with  finite differences. The scheme is second-order accurate in both space and time.  In the code, lengths are  normalized with respect to the full domain height, $L =$~4,000~km, and time is normalized with respect to the Alfv\'en crossing time computed as,
\begin{equation}
    \tau_{\rm A} = \int_0^L \frac{1}{v_{\rm A}}{\rm d}z, 
\end{equation}
which gives $\tau_{\rm A} \approx$~18~min in our model. This is the time for an Alfv\'enic perturbation to travel from the photosphere to the corona, or the other way around.

We consider a monochromatic driver located at the photospheric boundary that excites Alfv\'en waves with a fixed frequency, $f$. We prescribe $v_\perp$ at $z=0$ as,
\begin{equation}
    v_\perp =  \left[1 - \exp \left(-\frac{t^2}{\sigma^2}\right) \right] \sin\left(2\pi f t \right), \label{eq:driver}
\end{equation}
with $\sigma = 1/2f$. The first factor in Equation~(\ref{eq:driver}) corresponds to a smooth transient, which is included so that the periodic driver does not start abruptly from $t=0$. Conversely, $b_\perp$ is not prescribed at the photosphere. Prescribing both $b_\perp$ and $v_\perp$ would impose a particular superposition of $Z^\uparrow$ and $Z^\downarrow$ at the boundary, which would artificially affect the computation of the reflection coefficient. Instead, we use a extrapolated condition for $b_\perp$ using the two internal points adjacent to the boundary. Regarding the  conditions at the coronal boundary, $z=L$, we use a extrapolated condition for $v_\perp$ and impose that this boundary is perfectly transparent to the incoming waves from below, as in the stationary case. To achieve this last requirement, we set $Z^\downarrow = 0$ at $z=L$, so that  $v_\perp$ and  $b_\perp$ are related by,
\begin{equation}
b_\perp = - \sqrt{\mu\rho_{\rm c}}\,v_\perp,    
\end{equation}
where $\rho_{\rm c}$ denotes the density at the coronal boundary.

With the results of the temporal evolution, we compute the time-dependent upward and downward energy fluxes according to the expressions in Equation~(\ref{eq:fluxes}). Then, we use the values of those fluxes at the boundaries to compute the time-dependent versions of the reflectivity, transmissivity, and absorption coefficients, following the forms in Equations~(\ref{eq:reftran}) and (\ref{eq:abs}), with the temporally averaged fluxes being approximated by,
\begin{equation}
    \left< F^{\uparrow\downarrow} \right> \approx \frac{1}{t} \int_0^t F^{\uparrow\downarrow}(t')\, {\rm d}t'.
\end{equation}

First, we analyze the evolution of an Alfv\'en wave with a frequency of 5~mHz. A  spatial resolution of 250~m is used in this simulation.  Figure~\ref{fig:temporal5mHz} displays time-distance diagrams of the Els\"asser variables $Z^\uparrow$ and $Z^\downarrow$, which  correspond to the upward and downward components of the wave, respectively. During the first Alfv\'en crossing time, the propagation is predominantly in the upward direction until the wave reaches the sharp transition region located at $z/L \approx 0.55$. Then,  wave reflection drives perturbations in $Z^\downarrow$, which become visible in the time-distance diagram. After a brief transient, it is obvious that a stationary pattern is achieved very quickly. The simulation is run up to a maximum  time of $t=100\,\tau_{\rm A}$, although Figure~\ref{fig:temporal5mHz} only displays the results until $t=5\,\tau_{\rm A}$, as  the evolution does not show appreciable differences for larger times.

 \begin{figure}
   \centering
   \includegraphics[width=0.95\columnwidth]{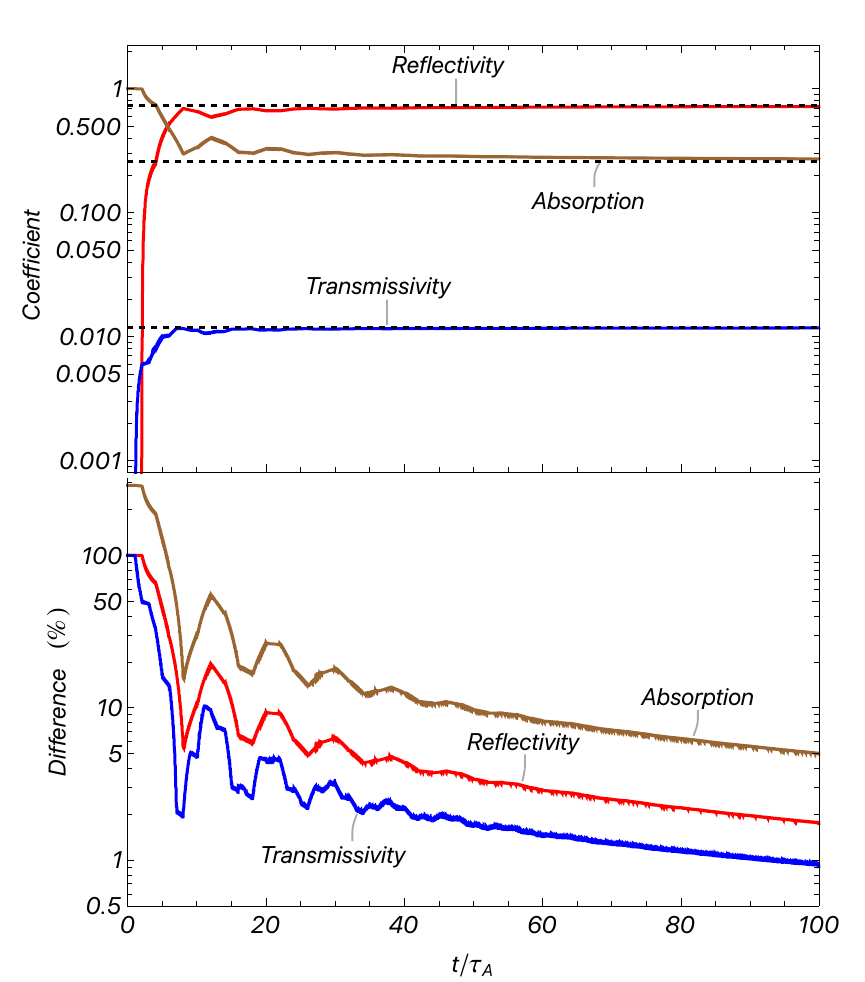}
      \caption{Time-dependent results for an Alfv\'en wave with $f=$~5~mHz. Top: Reflectivity, transmissivity, and absorption coefficients as functions of  time. The horizontal dashed lines denote the corresponding results in the stationary regime. Bottom: Percentage differences (in absolute value) between the time-dependent and the stationary coefficients as functions of  time. Time is normalized with respect to the Alfv\'en crossing time.}
         \label{fig:time5mHz}
   \end{figure}

The time-dependent reflection, transmission, and absorption coefficients have been computed for the 5~mHz simulation and are shown in Figure~\ref{fig:time5mHz}. The top panel illustrates the temporal evolution of these coefficients, with horizontal lines indicating the corresponding stationary values for comparison. As expected, the time-dependent coefficients consistently converge to their stationary counterparts as time progresses. To analyze this convergence, the bottom panel displays the percentage difference (in absolute value) between time-dependent and stationary results.  For all three coefficients, this difference diminishes over time, though the convergence rates vary among the coefficients. The transmissivity  converges the fastest, followed by the reflectivity, and finally, the absorption. By the end of the simulation, at $t=100\,\tau_{\rm A}$, the percentage differences from the stationary values are approximately 1\% for the transmissivity,  2\% for the reflectivity, and  5\% for the absorption. The slower convergence of the absorption coefficient can be partly attributed to its dependence on both the reflectivity and transmissivity, as its calculation inherently includes the percentage differences associated with these two coefficients.

 \begin{figure}
   \centering
   \includegraphics[width=0.95\columnwidth]{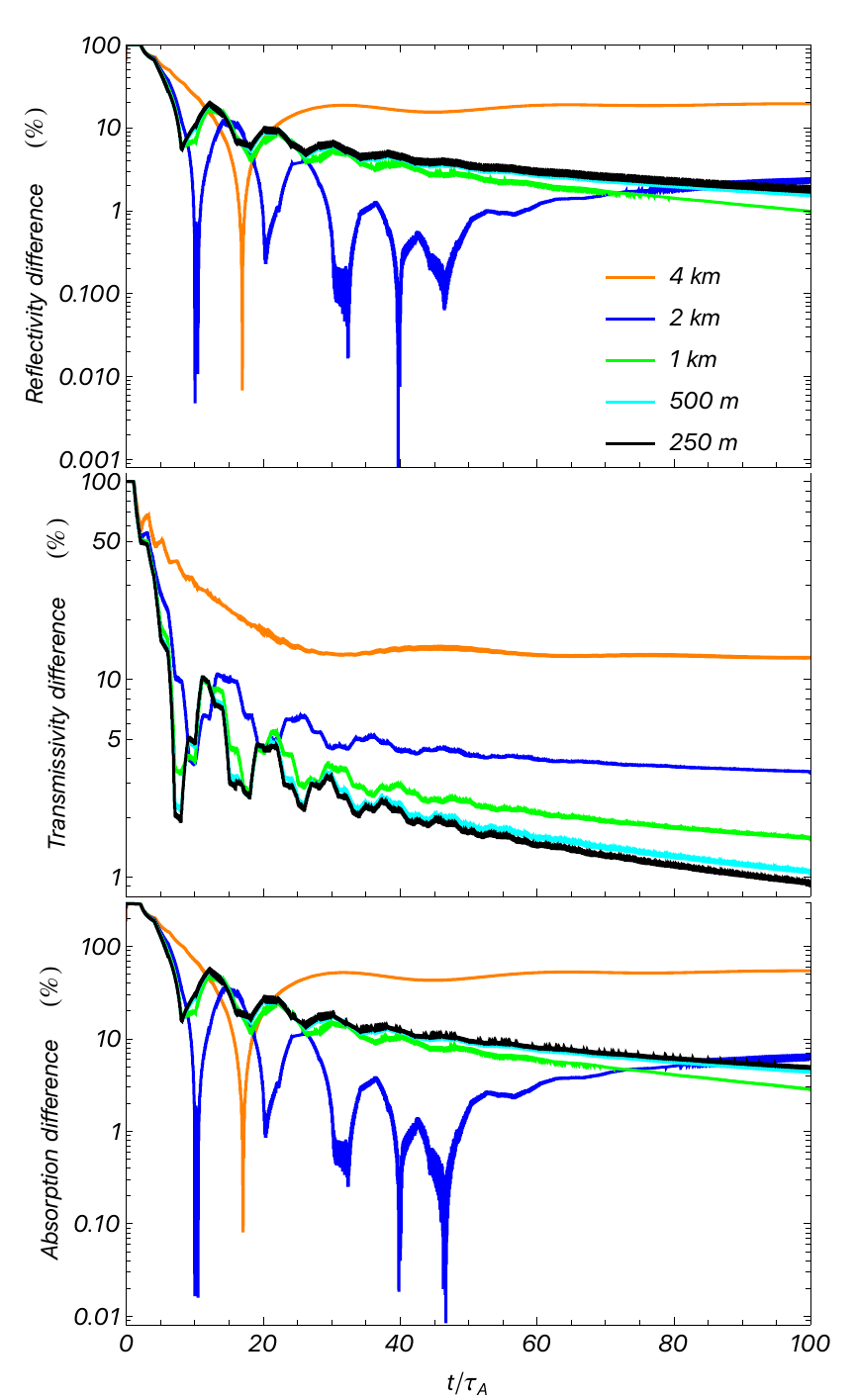}
      \caption{Time-dependent results for an Alfv\'en wave with $f=$~5~mHz. Percentage differences (in absolute value) between the time-dependent and the stationary reflectivity (top), transmissivity (mid), and absorption (bottom) coefficients as functions of  time. The various color lines correspond to different spatial resolutions. Time is normalized with respect to the Alfv\'en crossing time.}
         \label{fig:time5mHzreserror}
   \end{figure}

We have investigated the effect of spatial resolution on the convergence to the stationary regime. The simulation of the 5~mHz Alfv\'en wave  was repeated using coarser resolutions of 500~m, 1~km, 2~km, and 4~km. Subsequently, the percentage differences between the time-dependent and stationary coefficients were calculated. The results are presented in Figure~\ref{fig:time5mHzreserror}. The simulations with resolutions of 2~km and 4~km exhibit markedly different behavior compared to the others. In these cases, convergence to the stationary regime is not achieved, as the percentage differences for reflectivity and absorption increase at later times. Interestingly, at certain specific times, these differences appear to reach very small values. However, this is a result of the time-dependent coefficients crossing the stationary values due to the highly oscillatory behavior observed in these low-resolution simulations. These findings clearly show that the spatial resolutions of 2 km and 4 km are insufficient to accurately capture the evolution of the 5 mHz Alfv\'en wave. In contrast, a smooth convergence to the stationary regime is observed in simulations with resolutions of 1~km, 500~m, and 250~m. Among these, the convergence of the transmissivity exhibits a stronger dependence on spatial resolution, with finer resolutions leading to faster convergence. However, the reflectivity and absorption show no significant dependence on spatial resolution across these high-resolution simulations.

 \begin{figure}
   \centering
   \includegraphics[width=0.95\columnwidth]{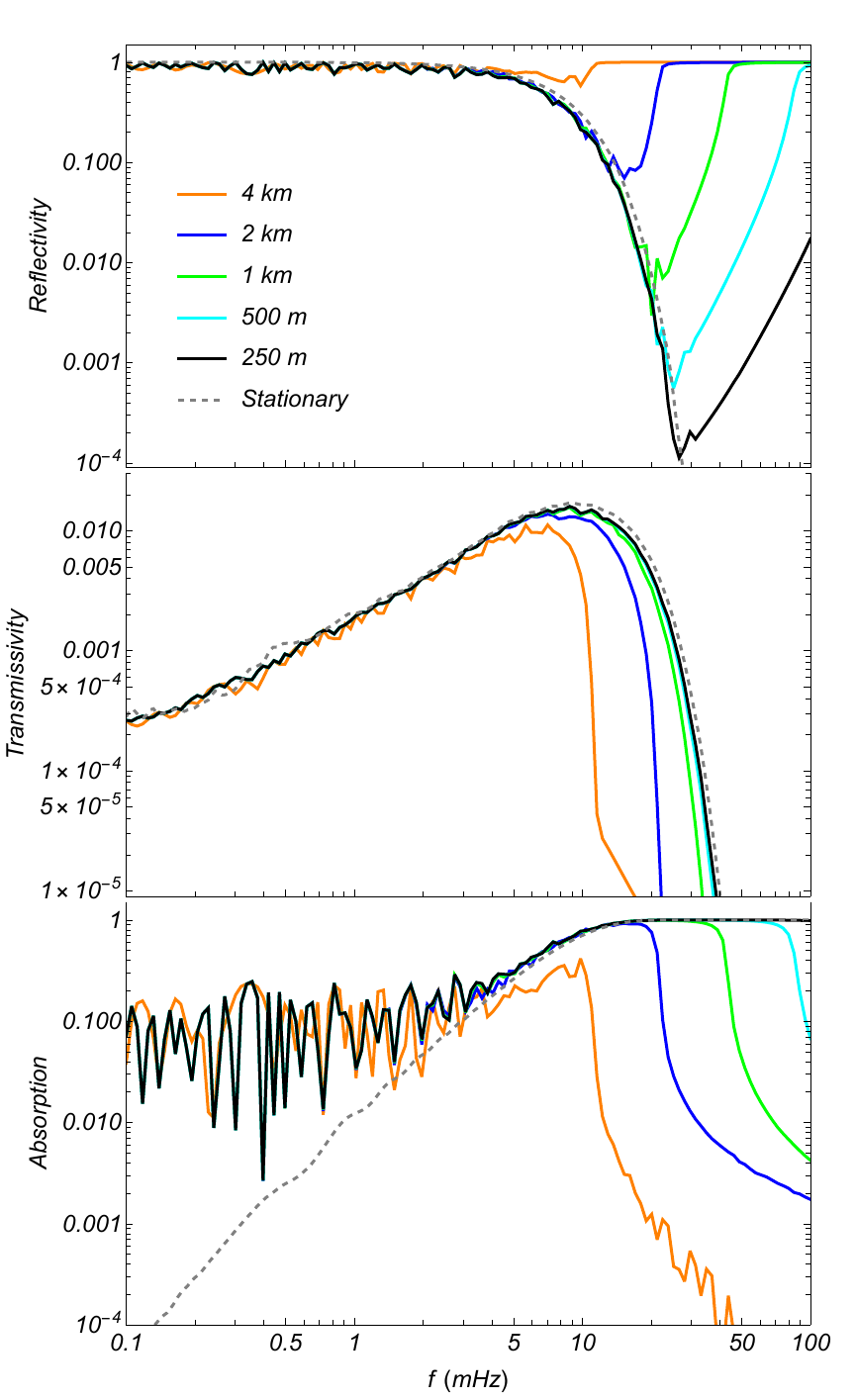}
      \caption{Time-dependent results. Reflectivity (top), transmissivity (mid), and absorption (bottom) coefficients as functions of the Alfv\'en wave frequency for $t=20\,\tau_{\rm A}$. The solid lines with various colors correspond to different spatial resolutions, while the gray dashed line denotes the  results in the stationary regime.  }
         \label{fig:freqscol}
   \end{figure}

 We now consider wave frequencies ranging from 0.1~mHz to 100~mHz,  as in the stationary computations of Section~\ref{sec:sta}, and investigate the convergence to the stationary regime across the entire frequency range. The frequency range has been discretized into 126 individual frequencies, spaced logarithmically. For each frequency, we performed five simulations at resolutions of 250~m, 500~m, 1~km, 2~km, and 4~km, resulting in a total of 630 simulations.  Figure~\ref{fig:freqscol} shows the reflectivity, transmissivity, and absorption coefficients obtained from the time-dependent simulations as functions of the wave frequency. The stationary coefficients are also plotted for comparison. Although all simulations were run up to a maximum time of $t=50\,\tau_{\rm A}$, the coefficients displayed in  Figure~\ref{fig:freqscol} were calculated  at  $t=20\,\tau_{\rm A}$. This time was chosen somewhat arbitrarily, as the results at later times show no significant differences for most frequencies, with the notable exception of the absorption coefficient, which is discussed later.

 \begin{figure}
   \centering
   \includegraphics[width=0.95\columnwidth]{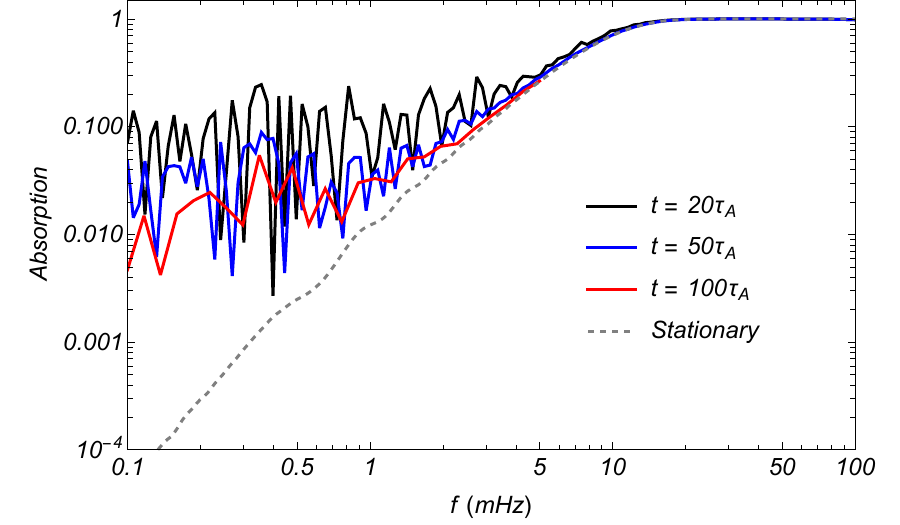}
      \caption{Time-dependent absorption coefficient as a function of the Alfv\'en wave frequency for $t=20\,\tau_{\rm A}$, $t=50\,\tau_{\rm A}$ and $t=100\,\tau_{\rm A}$. A spatial resolution of 250~m has been used. The gray dashed line denotes  the stationary result.  }
         \label{fig:abs}
   \end{figure}

Both reflectivity and transmissivity exhibit good convergence to the stationary results at low frequencies, where the results are independent of the spatial resolution. Conversely, as the frequency increases, the time-dependent coefficients begin to diverge from their stationary counterparts. The coarser the spatial resolution, the lower the frequency at which this divergence occurs. This marks a critical frequency beyond which the chosen resolution is no longer sufficient to resolve the spatial scales of the waves. At higher frequencies, we observe that the reflectivity is artificially enhanced, leading to a decrease in both transmissivity and absorption relative to the stationary values. For the finest resolution considered (250~m), this artificial increase in reflectivity appears at frequencies above approximately 30~mHz, whereas for the coarsest resolution (4~km), the critical frequency drops to about 4~mHz. These findings are consistent with the results discussed earlier in Figure~\ref{fig:time5mHzreserror} for the 5~mHz wave.

In contrast to reflectivity and transmissivity, the absorption does not exhibit good convergence to the stationary result at low frequencies. Figure~\ref{fig:abs}  compares the time-dependent absorption coefficient at $t = 20\,\tau_{\rm A}$ and $t = 50\,\tau_{\rm A}$ for a resolution of 250~m. Although the absorption at the later time shows some overall improvement in convergence, particularly at intermediate frequencies, it remains  larger than the stationary value in the low-frequency range. To assess whether convergence continues to improve at longer times, we extended some of the low-frequency simulations up to a final time of $t = 100\,\tau_{\rm A}$. Specifically,  we considered 26 logarithmically spaced frequencies between 0.1~mHz and 5~mHz for these extended runs. The corresponding absorption results at the final time are superimposed in Figure~\ref{fig:abs}. Once again, better convergence is obtained at intermediate frequencies, whereas for  $f \lesssim 1$~mHz the results remain inconclusive as to whether the stationary absorption value would eventually be reached at sufficiently long times.

Various factors may contribute to the poor convergence of the absorption at low frequencies. As previously discussed, the absorption is computed from the reflectivity and transmissivity, and thus it inherently accumulates the errors from these two coefficients. At low frequencies, where the transmissivity is negligible, even small errors in the calculation of the (dominant) reflectivity can lead to large relative errors in the much smaller absorption.  In fact, as shown in Figure~\ref{fig:freqscol}, the time-dependent reflectivity curve exhibits a noisy character at low frequencies, even for the finest  spatial resolution considered. This noise is likely related with the numerical approximation to the calculation of the temporally averaged energy fluxes at the photospheric boundary. It is also noteworthy that the stationary absorption is extremely small at low frequencies, owing to the weak physical dissipation. Consequently, the intrinsic numerical dissipation of the finite-difference scheme could have some influence in this frequency range. Nevertheless, the close agreement of the results across different spatial resolutions suggests that such numerical dissipation effects are unlikely to play a significant role.

\section{Conclusion}
\label{sec:conc}

We theoretically studied the propagation of pure Alfv\'en waves from the photosphere to the corona with the aim of comparing the results from the stationary approximation to those obtained by considering the full temporal evolution. For the stationary computations,we adopted the method used by \citet{soler2017,soler2019}, albeit with a simpler chromospheric model. In addition to the stationary calculations, we numerically solved the linearized MHD equations for the Alfv\'en wave perturbations using a time-dependent code. The results from both approaches were then compared.

 Considering a spatial resolution of 250~m in the time-dependent code, we found that the time-dependent  transmissivity progressively approaches its corresponding stationary value  in the entire frequency range as time increases. The convergence is relatively fast, with only a few Alfv\'en crossing times required for the stationary results to provide an excellent approximation. Concerning the reflectivity, we also found convergence toward the stationary values, but only for frequencies lower than 30~mHz. This critical frequency shifts to higher values as the spatial resolution increases, and to lower values as the resolution decreases.  In contrast,  the time-dependent results yield a higher absorption than the stationary results at low frequencies (below approximately 1~mHz), even after many Alfv\'en crossing times. This threshold frequency may vary depending on the strength of the background magnetic field. The discrepancy between the time-dependent and stationary absorption is likely explained by the fact that small relative errors in computing the reflectivity (the dominant coefficient at low frequencies) can lead to large relative errors in the estimation of the very small absorption.

The comparison of the results for different spatial resolutions shows that insufficient resolution leads to an artificially enhanced wave reflectivity, which in turn reduces the transmissivity and, at high frequencies, also the absorption. Even a spatial resolution as fine as 250~m is not entirely free from these issues within the considered frequency range, with noticeable effects on the reflectivity appearing at frequencies above approximately 30~mHz, as already discussed. While using even finer resolutions is feasible in the 1D simulations performed here, it could become a significant challenge in multi-dimensional simulations.

Recapitulating, only minor differences are observed between the stationary and time-dependent approaches regarding  reflectivity at high frequencies and absorption at low frequencies. These differences tend to gradually decrease as spatial resolution and simulation time increase, further reinforcing the validity of the stationary approximation.

Since the goal was to perform a large number of simulations covering a broad frequency range, we used a simplified 1D chromospheric model, which kept the computational costs within reasonable limits. As a consequence of this simplification, the phase mixing of Alfv\'en waves caused by cross-field gradients is not accounted for in this work  \cite[see, e.g.,][]{DeMoortel2000,Tsiklauri2001,Tsiklauri2002,Boocock2022a,Boocock2022b}. Previous studies have shown that phase mixing plays an important role in Alfv\'en wave dissipation in the lower chromosphere \citep{soler2019}. Capturing this effect would require multi-dimensional simulations. Another effect not included in this 1D model is the coupling between magnetoacoustic and Alfv\'en waves that takes place in more complex configurations \citep[see, e.g.,][for a comprehensive review]{callybogdan2024}.

A potentially important limitation of this study is the assumption of linearity. A natural next step is to extend the analysis to nonlinear simulations. In this context, the nonlinear coupling between Alfvén waves and slow magnetosonic waves may play a significant role \citep[see, e.g.,][]{antolin2010,arber2016,kuzma2020,ballester2020}.

\begin{acknowledgements}
      This publication is part of the R+D+i project PID2023-147708NB-I00, funded by MCIN/AEI/10.13039/501100011033 and by FEDER, EU. The author acknowledges the  student Llu\'is Nadal for performing some preliminary  computations and tests during his Master's project.  The author is also grateful to the anonymous referee for useful remarks.
\end{acknowledgements}

   \bibliographystyle{aa} 
   \bibliography{refs} 

\end{document}